\documentclass[english]{llncs}
\usepackage[T1]{fontenc}
\usepackage[latin9]{inputenc}
\usepackage{graphicx}

\usepackage{babel}

\title{Smart Auto Insurance: High Resolution, Dynamic, Privacy-Driven, Telematic Insurance}

\titlerunning{Title}
\author{
Michael Bartholic\inst{1} \and Zhengrong Gu \inst{1} \and Jianan Su \inst{1} \and Justin Goldstein \inst{1} \and Shin'ichiro Matsuo\inst{1}}
\institute{Georgetown University \\
\texttt{\{mwb70,zg120,js4488,jjg130,sm3377\}@georgetown.edu}}
\date{\today}

\begin{document}
\maketitle

\begin{abstract}
Data driven approaches to problem solving are\textemdash in many regards\textemdash the holy grail of evidence backed decision making. Using first-party empirical data to analyze behavior and establish predictions yields us the ability to base in-depth analyses on particular individuals and reduce our dependence on generalizations. Modern mobile and embedded devices provide a wealth of sensors and means for collecting and tracking individualized data. Applying these assets to the realm of insurance (which is a statistically backed endeavor at heart) is certainly nothing new; yet doing so in a way that is privacy-driven and secure has not been a central focus of implementers. Existing data-driven insurance technologies require a certain level of trust in the data tracking agency (i.e. insurer) to not misuse, mishandle, or over-collect user data. Smart contracts and blockchain technology provide us an opportunity to re-balance these systems such that the blockchain itself is a trusted agent which both insurers and the insured can confide in. We propose a "Smart Auto Insurance" system that minimizes data sharing while simultaneously providing quality-of-life improvements to both sides. Furthermore, we use a simple game theoretical argument to show that the clients using such a system are disincentivized from behaving adversarially.

\vspace{0.2cm}

\end{abstract}

\setcounter{secnumdepth}{3}

\section{Introduction}
\subsection{Background}
Auto insurance is traditionally inflexible, can often be expensive, and yet is usually a required part of vehicle ownership. The costs go towards paying for damages when accidents/incidents occur. Because of this, the cost of auto insurance is directly proportional to the risk.
Traditionally, this cost is assigned by the efforts of actuaries and statisticians based on risk tables of particular categories of individuals. However, this type of calculation must look at the average risk in the categories and thus does not necessarily reflect the particular behavior of a given individual.
Hence, in order to minimize costs to all parties, it may be desirable to create custom tailored auto insurance based on the behavior and risk level of a specific individual or set of individuals~\cite{CasualtyActuarialSociety}.
Recently a number of tools have been developed by insurance companies that are intended to measure a drivers behavior over a window of time and give them a rating based on performance of safe driving. For example: Progressive Snapshot. However, many individuals do not participate in such schemes due to privacy concerns or lack of clear benefit. Even if we assume insurance organizations are not intending to misuse or otherwise share the data they collect (perhaps not a viable assumption), it is unsafe to collect lots of user data in one centralized organization. 

\subsection{Related Works}
Our investigation of data-driven insurance is certainly not alone and there are a number of existing works in telematic auto insurance and other insurance domains. There are existing data-driven products by well established insurance companies such as Progressive's "Snapshot"~\cite{Snapshot:online} 
program and Allstate's "Drivewise"~\cite{Allstate:online} however these are primarily add-ons to existing insurance plans with periodic savings or adjustments. There are also a number of startups engaging with this domain, notably Arity~\cite{Arity:online} 
whose offerings focus on creating value for organizations with tracking based analytics. Most similar in premise is likely the "Smart Cyber Insurance" scheme proposed in~\cite{SmartInsurance}, where a ledger based approach with a searchable encryption scheme is used to dynamically rate risk levels for IoT devices. We investigate these works further in the Evaluation subsection. 

\subsection{Contributions}

Key contributions of this work revolve around the use of the blockchain as neutral trusted platform in the data collection and computational process. We propose a practical scheme using commonly utilized technologies in the space such a smartphone and on-board diagnostics (OBD-II) reader~\cite{OBD-II}\cite{10.1007/978-3-319-07776-5_43}\cite{8697833} based monitoring.

The advancement here is not in the detection of such behaviors or incidents, but instead in the way the data is protected, stored, and utilized securely (while possibly producing incentive to act more safely). The use of blockchain makes it easier to verify and process insurance claims, as well as securely and privately collect data from different stakeholders. We introduce an effective means to do, so called, "usage-based insurance" securely in a variety of novel use cases.

Both dynamic pricing and offering service to only certain clients are possible given this kind of data scheme. Dynamic pricing of insurance premiums follows more closely along the lines of the Smart Cyber Insurance system conceived in ~\cite{SmartInsurance}. However, only offering service to individuals below a certain risk level is a behavior that is already practiced by some companies which explicitly market as "service for safe drivers". Along with usage-based pricing, these are much more interesting use cases to explore because they involves less of a paradigm shift for existing companies. It has been seen that current insurance companies are reluctant to implement something truly dynamic because of implementation difficulties and regulatory concerns~\cite{SmartInsurance}.

\section{Data Driven Auto Insurance}
\subsection{Overview}
A number of tools have been developed to measure divers' behavior and give a score for their drivers' performance. Plug-in devices or mobile applications do not punish drivers but only offer discounts for \textit{good} behaviors.
These devices are heavy trackers and generally involve streaming detailed, high resolution data about drivers' personal activities to the insurance company for analysis. Instead, it is safer to only collect what you need and limit high resolution private data to avoid compromising the user. 
Collecting centralized data is risky because it enables central points of failure. Regulation on data management does not reduce the inherent risk in collecting lots of user data in a single location. When there is a single resource that is being trusted to maintain the privacy of information, you introduce a situation where the impact of compromising them is highly valuable, therefore dramatically increasing risk of compromise.

\subsection{Challenges}

Insurance operates as a business and is therefore driven primarily by cost and profit considerations. The major difficulties with the existing business includes high risk of large data disclosure, limited ability of data tracking, low transparency of assessment system. The auto insurance industry as a whole experiences a number of common challenges: processing insurance claims is a massively slow and resource intensive process which involves determining fault assessing values, and negotiating between insurance companies as well as individual clients.

More specific to data-driven or "telematic" insurance schemes, challenges include promoting mass participation, user privacy, and demonstrating clear advantages to users. While there is precedent for their use, participation in existing data-driven systems such as Progressive Snapshot~\cite{Snapshot:online} and Allstate Drivewise~\cite{Allstate:online} is not a widespread practice. These offerings tend to offload data directly to the organizations with little or no user control over the extent in which their data is used. Clearly these existing technologies are not particularly privacy focused. Likewise, is it scarcely obvious to the user exactly what risk assessment schemes are being employed. In this way these systems are somewhat of a "blackbox" with an unclear value proposition as mere add-ons to traditional schemes.

\subsubsection{Soundness}

A fair data-driven insurance system needs to be dynamically responsive to user behavior, but also verifiable when necessary and consistent in assigning risk. Traditional insurance offerings leave the client with little ability to ensure that the decisions about their insurance fee or claims are being made in good faith or actually utilizing the data they are providing. This can lead to inconsistency and arbitrary decision making that is motivated by causes outside of the user's immediate circumstances. Insurance organizations might make decisions based on profit margins, quotas, or a lack of sufficient evidence. 

\subsubsection{Privacy}

Existing data-driven auto insurance schemes are notably not privacy focused. These systems encourage the installation of tracking equipment and applications with little guarantee in the way of data use beyond suggestions of potential monetary savings. Instead of simply transmitting user data off for external computation and data-mining, we believe an ideal embodiment of a data-driven insurance product would perform computations locally and only transmit/disclose what is critical to the insurance process. While the existing paradigm is for insurance organizations to process data themselves, it should be possible to move this processing to the user side and limit unnecessary data sharing. In such a system the user's device could process the risk itself\textemdash according to predefined rules\textemdash and publish risk scores along with related info such as what specific rules are violated and evidence that was used to make the assessment. As specific violated rules and evidence would be of secondary importance compared to the risk assessment itself, these ancillary outputs could be protected and released in a controlled manner, only when necessary.

\subsection{Usage-Based Pricing Structure}
An optimal embodiment of data-driven insurance would be one where expenditures on either side of the transaction are minimized with the resolution of input data directly corresponding to the resolution of costs and fees associated with participation in the system~\cite{NAIC:online}. A system where risks and costs can be assessed actively with minute to minute or second to second precision lends itself to a novel pricing where the expense to the user can be directly associated with their use of the insured vehicle. This usage-based pricing structure is something we believe can be embodied by our proposed system. Such usage-based pricing can also introduce new use cases and product offerings that are presently unattainable by traditional pricing and data collection schemes.

\section{Our Solution: Smart Auto Insurance (System Design)}

Our solution is based on the characteristics of behaviors and events from the view of the insurance company. Behaviors and events are obfuscated from the view of the insurance company as to not reveal precise sensitive details about the insured party. Data is logged with risk scores analyzed and published actively while the specific risk-contributing behaviors and data logs are encrypted and not shared in real time. This time buffer reduces compromising insight into each user's personal behavior. In order to measure and track risk levels, rules are predefined by the insurance organization to weight different behaviors and data points.

Decryption only takes place for particular instances of data with the permission of the insured party. By protecting information and making it easily accessible when needed (making a claim, etc) we reduce the difficulty of making a claim when complicated negotiations need claims and faults need proof.

Existing insurance offerings require large amounts of statistical background about the habits of groups of individuals. We envision several novel use cases beyond traditional coverage of individuals. First, an insurance product for organizations: for fleets of cars in a set of company owned vehicles, rental companies with short term coverage, or offered to contractors for fleet organizations such as Uber(Eats), Postmates, etc. Second, an insurance product for individuals with emphasis on usage-based dynamic pricing. Third, an insurance product for robots/autonomous vehicles:
measuring the risk of a vehicle's use directly introduces the possibility to insure autonomous or remotely operated vehicles based on corresponding risk factors.

\subsection{Stakeholders}
The system is primarily constructed of insured parties and insurance companies. In traditional insurance it is particular individuals who are insured in their use of a given item (i.e. a specific person's use of a vehicle), as different individuals tend to have different risk profiles that affect their likelihood of filing a claim (i.e teenage boys are generally riskier than teenage girls). While this type of fee structure can certainly still be supported, here we put specific focus directly on the items that are insured because we are more intimately interested in the risk profiles of the items themselves and how they are used.

\subsection{System Model}
The Smart Auto Insurance system is comprised of 1 to N insurance organizations and any number of individuals enrolled with the organization of their choice. Their system can theoretically be entirely separate between competitor organizations, however we will see that there can be certain advantages to operating with a combined, interoperable system.

Enrollment in the Smart Auto Insurance system identifies a vehicle with the insurance provider alone. With of this identification, the system may be considered as an addition to traditional schemes. Smart Auto Insurance allows for factoring in policy type information associated with the account that would not necessarily be conveyed by other data measures (vehicle type, age, etc). We emphasize that being able to utilize high resolution first-person data may reduce the need to resort to broader actuarial predictions or generalizations. We envision the Smart Auto Insurance system as a product of its own. In this way, the scheme can be much more personalized to individual behaviors and not necessarily need a possibly biased "client profile," since risks can be measured more directly~\cite{TelematicsGenderDiscrimination}.

Upon enrollment in the system, each vehicle will be given an OBD-II or similar diagnostics processing device or an individual may install a smartphone application. These diagnostics tools will allow for the logging and processing of vehicle data to generate EventData (risk values, violated rules, and evidence).

Most notably, metrics are not transmitted directly. Risk rules are computed and then results are published as EventData according to its specific use. Since the insurance fee can be determined without direct data access, it is possible that there is no need to reveal the data unless someone is trying to make an insurance claim or otherwise audit the system.

\begin{itemize}
    \item Parties
    \begin{itemize}
        \item Insured Vehicles, Owners of
        \item Insurance Companies
    \end{itemize}
    \item Assumptions
    \begin{itemize}
        \item Insured parties seek to insure specific items/vehicles/etc.
        \item Insurance risk rating rules can be predefined
        \item Existence of Blockchain Hosts
        \item Existence of the Cryptography as a Service System (4.1.2)
        \item (Optionally) Existence of global data sources via an oracle of that information
    \end{itemize}
\end{itemize}

\begin{figure}[t]
\begin{center}
\includegraphics[scale=0.22]{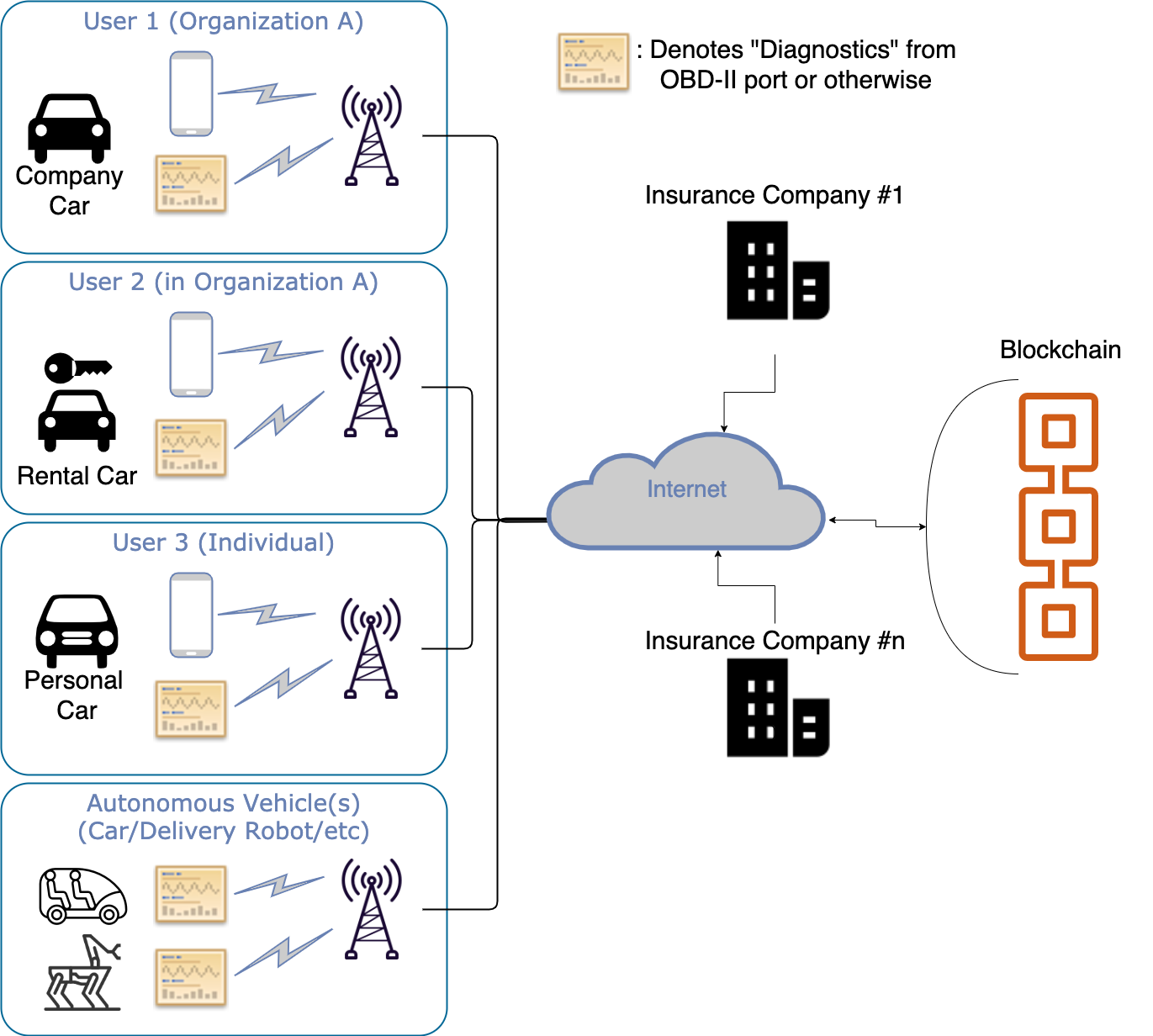}
\caption{System model of Smart Auto Insurance}
\label{fig1}
\end{center}
\end{figure}

\subsection{Primary Processes}

The system can be summarized by four high level procedures split between local and smart contract based execution. These phases include collecting the requisite data, securely processing it according to predefined rules, generating disparate outputs for publishing, and publishing these outputs according to their use and access limitations within the system. Procedures include:

\begin{description}

\item[Log Data:]
Active monitoring for vehicle behaviors and incidents, logging data locally (off-chain) according to observations for later computation.

\item[Risk Assessment:]
Periodic assessment of risk levels and "rule violations" locally (off-chain) utilizing the data that is recorded in previous procedure. Computed using rule table published by insurer ahead of time.

\item[Publish Events:]
After RiskAssessment, publish computed risk levels; publish (1) violated rules, and protect this information for controlled release; publish (2) "evidence data" that was used to trigger the given rule, and protect this until joint action with user and insurer.

\item[Reveal Protected Information:]
Decrypt the encrypted (1) violated rules or (2) evidence data using the single use symmetric key that encrypted it.

\end{description}

\section{Proposed System in Detail}

Individuals have an incentive to participate in the system for the possibility of saving money and they do not have to reveal personal information. Likewise, insurance organizations have an incentive to participate in the system for the claim processing ease and possible cost savings.
Both parties have an incentive to unlock personal information when an insurance claim is being made. 
If risks are assessed correctly, this also creates a pressure to drive more safely. The system permits a variety of possible scoring mechanisms as determined by implementers, which can lead to varied individual incentives and possible market competition.

\subsection{Privacy Protection}

For our design of Smart Auto Insurance, we strive for the highest privacy privacy that is practically attainable. Under ideal circumstances would we not be sharing any information, but in order to construct a practical system there needs to be some concessions. We are interested mainly in eliminating the ability for arbitrary use of large amounts of personal information while still supporting the required computations in throughout the system.

\subsubsection{Privacy Goals}
\rule[-10pt]{0pt}{10pt}\\
Therefore, our privacy goals are as follows:
\begin{enumerate}
    \item Eliminate the need for offloading raw data to external parties.
    \item Eliminate the need for black box data processing.
    \item Do not publicly publish potentially compromising data.
    \item Do not trust any single party with protecting potentially compromising data.
    \item Reveal only the information required to perform a given insurance related function (determining payment amounts, processing claims, etc.).
    \item Obfuscate the origin of the information that must be shared.
\end{enumerate}

\subsubsection{Cryptography as a Service with Evolving-Committee PSS}

A key facet of the privacy protection in the Smart Auto Insurance system relies on the ability to trust a blockchain itself with controlled release of secrets. In this way, the system protects data for privacy purposes while operating as an independent platform that yields information when the right conditions are met.

The work by F. Benhamouda et al. describes a robust system for a "Evolving-Committee Proactive Secret Sharing" scheme which introduces the ability for arbitrary secret keeping in a resilient manner~\cite{Craig2020}. The full applications of such a scheme are beyond the scope of this paper, but our interest for the Smart Auto Insurance system is in how it can enable, as they describe, "Cryptography as a Service". 

Cryptography as a Service in this circumstance is effectively a tool that allows for the Smart Auto Insurance system to encrypt data, hand off the key, and trust that the key will be released in a controlled fashion. The control conditions can be defined by either a time duration or join action between parties. Once released, the Smart Auto Insurance's RevealProtectedInformation procedure can access the key, download the relevant data (either a violated rules list or evidence data), and decrypt the information for further use. Critically, this scheme allows for the protection of arbitrary secrets and gives us a means to dynamically protect individual EventData publications based on their privacy requirements without trusting any single party. 

\subsubsection{Data Management}

Offline computation requires that insurance organizations commit to specific (versionable) breakdowns of risk rules, which can make analysis less hidden and increase transparency. The actual scoring is somewhat arbitrary but would need to be computable offline, from locally recorded measurements. There should be joint action from both the user and the company for revealing evidence data, which is stored data off-chain in a secure manner. A possible inclusion to the system could allow users to include further evidence than what is recorded automatically when such as photos and video from a dash-cam, smartphone, or other device. 

"EventData" is the collection of computation inputs and results that are produced and published every time the RiskAssessment procedure is run. EventData is composed of three distinct components: a summed risk score, the list of violated rules that were triggered to produce the risk score, and the evidence data that was processed to trigger these rules. Each of these components has a different use in the system and protection methodology as described below.

\begin{itemize}
    \item Off-chain Data
    \begin{itemize}
        \item Logged data: Data logs are recorded periodically to local storage on the insured item's device for processing by the RiskAssessment procedure.
        \item Device signing key: Upon enrollment with their respective insurance organization, a signing key is generated for local storage that will be used to validate that data contributions originate from the device in question. 
        \item External data: (Optional) The system can permit the inclusion of rules based on data that originates from an oracle of the information. For example, time-based rules with an external time authority seem like an obvious interest; however, risk assessments based on public
        records of firms such as trulia.com\footnote{As a part of the service on the website trulia.com, crime maps are integrated from organizations such as spotcrime.com that aggregate crime reports from local agencies in various areas. It is not to say that such reports are unbiased or a perfect characterization of risk, but if deemed trustworthy, the conclusions of such organizations may be relevant in evaluating situational risk.} may also be of interest if they can be trusted.
        \item Evidence data: The third component of EventData, evidence data is encrypted with a single use symmetric key and then published every time the RiskAssessment procedure is run. This data is used to back insurance claims as well as validate that the risk assessment is accurate in the event of an audit. Due to the weight of the data (possibly upwards of megabytes per publication), it is stored off-chain with the URL to the encrypted data included on-chain. The precise off-chain location is not particularly significant, but if a decentralized (and incentivized) storage approach is desirable, then an integration with IPFS~\cite{IPFS} and/or Filecoin~\cite{Filecoin} is certainly attainable. See Table 2 for some examples.
    \end{itemize}
    \item On-chain Data
    \begin{itemize}
        \item User Device UUID: Each insured item/vehicle will have a unique identifier registered with the respective insurance organization upon enrollment. This ID is known by the item's application itself as well as the insurance organization but obfuscated upon event publication to reduce tracking ability of other individuals.
        \item Rule tables: Insurance organizations will publish rule tables that describe how the RiskAssessment procedure should be evaluating the risk of various data points. Similarly to user contributions, insurance organizations will sign rule table publications to confirm their validity. As rules tables are identifiable by organization and version tracked on the ledger, the system will be able to identify what the applicable rule set is for a given time interval. See Table 1 for some examples.
        \item Event risk scores: The first component of EventData, the risk score evaluations from each run of RiskAssessment are published in plaintext for the insurance organization to use when billing the user.
        \item Violated rules lists: The second component of EventData, the list of violated rules is encrypted with a single use symmetric key.
        \item Encryption key for violated rules list: For each publication following the data processing by the RiskAssessment procedure, a single use symmetric encryption key (i.e. AES~\cite{2001}) is generated to encrypt just the list of violated rules. This key is then handed off to the Cryptography as a Service mechanism for protection of the secret and controlled release (see 4.1.2).
        \item Encryption key for evidence data: For each publication following the data processing by the RiskAssessment procedure, a single use symmetric encryption key is generated to encrypt just the evidence data that backs the RiskAssessment results. Likewise, this key is then handed off to the Cryptography as a Service mechanism for protection of the secret and controlled release (see 4.1.2).
    \end{itemize}
\end{itemize}

\subsection{Security and Soundness}
\begin{description}
\item[Against a malicious user:]

Under most circumstances we would have a strict requirement for ensuring that a user's data inputs are valid. Aside from trusted inputs provided by an oracle, we would be pressed to validate that user inputs are honest and correct. However, because of the transactional nature of such an insurance scheme and the requirement of proof when making a claim, we effectively create a game where users are disincentivized from behaving maliciously. The game is constructed as follows:

\begin{enumerate}
    \item The two parties are insured individuals and insurance organizations.
    \item Both parties seek to minimize costs and maximize self-benefit.
    \item Insured individuals will be required to pay an insurance fee for participation in the system at all times, not just when there is a claim.
    \item The fee required from individuals is proportional to the assessed risk of their behavior based upon their data inputs.
    \item Insured individuals will file claims to reduce their financial liability when certain events (i.e. damages) occur.
    \item Insurance organizations only accept claims which are accompanied by substantiated evidence demonstrating the circumstances of the damages.
    \item A user only receives a benefit from participating when they can make successful claims.
    \item Events which require a claim are unpredictable.
\end{enumerate}

Statements 2 and 4 imply that a user may have an incentive to provide false inputs to the system that do not accurately reflect their risk, in order to minimize costs. However, statement 8 implies that a user cannot easily know one way or the other whether an event requiring a claim will occur and adjust their behavior accordingly. Statements 5 and 6 imply that a user must be behaving honestly in order to make a valid claim and receive a benefit from the system. Statements 2 and 7 imply that a user would not participate in the system if they unable to make successful claims (and receive a benefit). As such a user couldn't choose to behavior honestly just to make a valid claim without behaving that way overall. In this way, a malicious user has weak chances of any gain from the system and more so little incentive to participate in the system at all due to the cost commitments of participating.

\item[Against a malicious insurance organization:]

Instead of the traditional actuarial process that happens behind closed doors following whatever rules and principles that the organization sees fit, by requiring that risk be computed in real time and off-line, we introduce the need for insurance organizations to predefine rules and publish clearly how risk is computed. This operating principle creates a number of distinct advantages for the security and soundness of the system. Most significantly, insurance organizations that seek to participate in such a system will be required to have a certain level of transparency is their rule definitions. Furthermore, this also decreases the insurance organization's ability to misuse the wealth of data it is being provided. In contrast to existing data driven schemes that simply feed data off for to remote servers for computation, Smart Auto Insurance increases the security and expectation of soundness experienced by the users by reducing an insurer's ability to leak, sell, or otherwise distribute user data en mass.

\end{description}

\subsection{Procedure Detail}

\begin{description}

\item[Log Data:] (Track behaviors/incidents)
This procedure is effectively a measurement loop which samples values each at its predetermined interval. Not every data point needs to be collected at the same frequency. In addition to measurement tracking there ought to be logging to detect connecting, removing the device itself, starting, and stopping use of the vehicle.
\begin{itemize}
    \item \textbf{Input:} list of points to measure and their frequencies
    \item \textbf{Output:} log files to be processed by RiskAssessment
\end{itemize}

\item[Risk Assessment:]
At an interval prescribed by the system's parameters, the logging device or mobile device runs the RiskAssessment procedure. This procedure utilizes the data logs and predefined rule-sets to determine the device's score in a given interval. Importantly, this process occurs locally on the measurement device using a rule-set that the insurer has published and signed.
\begin{itemize}
    \item \textbf{Input:} logged data (local), predefined rule-set (on-chain) 
    \item \textbf{Output:} $EventData$ comprised of the 1) risk score, 2) list of violated rules, 3) logged evidence data from the interval
\end{itemize}

\item[Publish Events:]
This procedure protects EventData and then publishes EventData to the ledger and external database. It generates single use AES symmetric encryption keys which are used to separately encrypt the violated rules and evidence. It then hands off these secrets for management by the Cryptography as a Service scheme. The mobile device doing the processing also signs its data contributions with the key registered with the insurer so that it is possible to identify the origin of the contribution if it ever needs to be decrypted. Finally the procedure invokes smart contracts to publish these user contributions (risk score, encrypted rules, and encrypted evidence). 

\begin{itemize}
    \item \textbf{Input:} All EventData (outputs from RiskAssessment)
    \item \textbf{Output:} Risk score (on-chain), encrypted list of violated rules (on-chain), single use symmetric key for violated rules, URL to encrypted off-chain evidence data, single use symmetric key for evidence
\end{itemize}

\item[Reveal Protected Information:]
This procedure runs via a smart contract to access and decrypt either or both the list of violated rules and the evidence of each offense. Following the use of Cryptography as a Service, (see section 4.1.2) when a time limit has passed or there is joint action between the client and insurer, we use the secret to decrypt and return plaintext data. This data can then be used to process an insurance claim, audit the system, etc. As violated rules and evidence are protected separately, they can be revealed separately: i.e violated rules revealed after a sufficient time as passed to reduce trackability of user; evidence, only upon making a claim.

\begin{itemize}
    \item \textbf{Input:} single use symmetric keys from Cryptography as a Service, violated rules ciphertext (on-chain), URL to evidence data
    \item \textbf{Output:} violated rules plaintext or evidence plaintext
\end{itemize}

\end{description}

\begin{table}[t]
\begin{center}
\resizebox{\textwidth}{!}{\begin{tabular}{|p{2cm}|p{1cm}|p{1cm}|p{1cm}|p{1cm}|p{1.5cm}|}
\hline

\multicolumn{1}{|c|}{\textbf{Organization ID}} & 
\multicolumn{1}{|c|}{\textbf{Rule \#}} & 
\multicolumn{1}{|c|}{\textbf{Field name}}& 
\multicolumn{1}{|c|}{\textbf{Data Type}}& 
\multicolumn{1}{|c|}{\textbf{Logical Rule}}& 
\multicolumn{1}{|c|}{\textbf{Output Penalty}} \\ \hline\hline

CompanyOne &
R1 &
Precipitation &
Boolean &
\mbox{$value$ is True} &
$+$5
\\
\hline

CompanyOne &
R2 &
Velocity &
\mbox{Float \(m/s\)} &
\mbox{$value \geq 30 $} &
\mbox{$+10\times$($value-30$)}
\\
\hline

CompanyOne &
R3 &
Acceleration &
\mbox{Float \(m/s^2\)} &
\mbox{$|value| \geq$ 1.34} &
$+$15
\\
\hline

CompanyOne &
R4 &
EngineRPM &
Integer &
\mbox{$value \geq 6000$} &
$+$10
\\
\hline

\end{tabular}}
\end{center}
\caption{Examples of rule table information}
    \label{tab:ruleData}   
\end{table}

\begin{table}[t]
\begin{center}
\resizebox{\textwidth}{!}{\begin{tabular}{|p{2cm}|p{1cm}|p{1cm}|p{1cm}|p{1cm}|p{1.5cm}|}
\hline

\multicolumn{1}{|c|}{\textbf{Field Name}} & 
\multicolumn{1}{|c|}{\textbf{Data Type}} & 
\multicolumn{1}{|c|}{\textbf{Sampling Freq.}} & 
\multicolumn{1}{|c|}{\textbf{Output Value}} &
\multicolumn{1}{|c|}{\textbf{Time recorded ISO8601}} \\ \hline\hline

Precipitation &
Boolean &
0.5 Hz &
True &
\mbox{'2021-01-31T16:40:44.26'}
\\

\hline
Precipitation &
Boolean &
0.5 Hz &
True &
\mbox{'2021-01-31T16:40:46.26'}
\\
\hline

Precipitation &
Boolean &
0.5 Hz &
False &
\mbox{'2021-01-31T16:40:48.26'}
\\
\hline

Velocity &
\mbox{Float \(m/s\)} &
2 Hz &
28 &
\mbox{'2021-01-31T16:40:47.26'}
\\
\hline

Velocity &
\mbox{Float \(m/s\)} &
2 Hz &
30 &
\mbox{'2021-01-31T16:40:47.76'}
\\
\hline

Velocity &
\mbox{Float \(m/s\)} &
2 Hz &
34 &
\mbox{'2021-01-31T16:40:48.26'}
\\
\hline

Acceleration &
\mbox{Float \(m/s^2\)} &
5 Hz &
2 &
\mbox{'2021-01-31T16:40:47.86'}
\\
\hline

Acceleration &
\mbox{Float \(m/s^2\)} &
5 Hz &
1.6 &
\mbox{'2021-01-31T16:40:48.06'}
\\
\hline

Acceleration &
\mbox{Float \(m/s^2\)} &
5 Hz &
0.8 &
\mbox{'2021-01-31T16:40:48.26'}
\\
\hline

EngineRPM &
\mbox{Integer} &
5 Hz &
6500 &
\mbox{'2021-01-31T16:40:47.86'}
\\
\hline

EngineRPM &
\mbox{Integer} &
5 Hz &
4000 &
\mbox{'2021-01-31T16:40:48.06'}
\\
\hline

EngineRPM &
\mbox{Integer} &
5 Hz &
2000 &
\mbox{'2021-01-31T16:40:48.26'}
\\
\hline

\end{tabular}}
\end{center}
\caption{Examples of log data}
    \label{tab:logData}   
\end{table}

\section{Implementation: Proof of Concept}

To demonstrate the viability of the system proposed here, a simple proof-of-concept was implemented using Java and Ethereum (Solidity). This implementation simulates the processes of logging data; accessing the data logs to run a RiskAssessment; collecting, encrypting, and publishing results; and finally retrieving encrypted data. To conduct the experimental we ran the simulation process described above 100,000 times while timing each phase and collected statistics about the mean, maximum, minimum, and standard deviation of run-times for each stage. The log files for each test represent a 5 minutes (300 second) interval with the frequencies denoted in table 1. Log files are pseudo-randomly generated in the LogData phase with weights that produce data that are qualitatively similar to their practical values. Table 1 denotes examples of the types of rules that could be computed by the RiskAssessment procedure. Table 2 demonstrates what data outputs of the system look like.

\section{Evaluation}

Finally, we evaluate the success of our proposed Smart Auto Insurance system against the identified weaknesses of traditional insurance, our privacy goals, and the limitations of the similarly motivated work Smart Cyber Insurance~\cite{SmartInsurance}.

Compared to the status quo, we believe Smart Auto Insurance represents an improvement to both privacy and data handling. By taking advantage of readily available data sources (OBD-II/diagnostics/etc) and well defined risks, we are confident that existing technologies can support the novel aspects of the system. By establishing methodologies for trusted processing of data on the user side and conveying the minimum information for billing, we adequately protect user information and only reveal protected details when the user makes a claim.

Both our system and the Smart Cyber Insurance system heavily emphasize the utility of dynamic score of risk for particular environments with the goal of creating a dynamic risk scoring system. A dynamic risk scoring system enables advanced pricing of insurance schemes. Both schemes measure the level of risk in an environment based on observed risk factors and both have sets of risk factors that can be assigned varying levels of severity.

A distinction between these systems is the nature their risks. Smart Cyber Insurance has an extremely broad set of possible risks that require constant updates from a trusted data source in order to remain viable. Smart Auto Insurance has a rather well defined and closed class of risk factors. It may be possible to exhaustively prescribe risk levels for relevant factors without routine updating. 

Furthermore, Smart Cyber Insurance has what can be described as a data input problem. In order to operate, the system has to presuppose some data input through the "device manager" and the administrator of the network. This may be a significant assumption because there are not necessarily simple and well defined ways to ensure comprehensive data tracking and input. More so, the collection of these data by the manager creates a risky data-pool. Ideally, a network administrator would know what is installed and would control it, but this can be a point of difficulty in any practical implementation. With Smart Auto Insurance, insurance offerings have already demonstrated effective data collection through of OBD-II readers and smartphones.

In Smart Cyber Insurance, the scheme is only specifically described in terms of a network of devices controlled by a single organization. This is not directly applicable to individuals and the individuals who participate under the organization are not in control of their data contributions. One of the main issues is that the data input methodologies that are most likely feasible for organizations do not necessarily work for individuals. Additionally, there is greater risk in specifically pairing risk levels to individuals in cyber insurance because you can create inadvertent targets for attacks. That is, in order to assign risks one needs to match vulnerabilities to devices, which can be compromising to individuals. The scheme works in collections of devices to reduce this exposure.

In Smart Auto Insurance, there is not the same concern for creating targets as there is for Smart Cyber Insurance because the nature of the risk is different. For auto insurance, the risk is almost exclusively the danger of monetary loss for the driver, company, and other drivers through operation of a vehicle. Furthermore, the Smart Auto Insurance risk is circumstantial or transient and not necessarily a trait which increases an individual's chance of being a target. Due to this, knowing of one's "unsafe" behavior in vehicle operation is unlikely to create a scenario that makes them a greater target for further risk in the way having many software vulnerabilities would in the realm of cyber insurance.

\subsection{Effectiveness}

\subsubsection{Privacy}
The main concern of the Smart Auto Insurance system is elevating and emphasizing privacy in the (data-driven) insurance process. Notably, the system succeeds in minimizing data sharing and the resulting privacy risks.

\subsubsection{Security and Soundness}
\begin{description}
\item[Against a malicious user:]

We find the system to be at minimal risk of malicious users gaining any practical benefit. Our analysis suggests that under typical operating circumstances a user would be unconfident in their ability to make dishonest successful insurance claims. Without the ability to make insurance claims to receive a benefit, malicious users have no incentive to participate in the system as participating has an associated monetary cost.

\item[Against a malicious insurance organization:]

While the nature of insurance claims makes it difficult to fully remove an insurance organization's ability to make arbitrary, closed-door decisions, we believe the Smart Auto Insurance system represents an improvement on the status quo by requiring the pre-definition of evaluation rules. Furthermore, by not sending data directly to the insurance organization and processing data locally, we achieve a system that dramatically reduces the ability for an insurer to mishandle user data.

\end{description}

\subsection{Efficiency}

To demonstrate the system is both feasible and practical, it is important to show that it is efficient enough to run in real time without disproportionately large computational resources. Efficiency of the secret storage, protection, and management is demonstrated by~\cite{Craig2020}, where it is shown that complexity follows on the order of the relatively small committee size with each party's computation on the order of log(N total parties). Therefore, our main concerns are with the performance of the logging, risk analysis, and encryption processes. 

\subsubsection{Space}

Space requirements of the system are not of particular concern because there is precedent for this type of data logging procedures. Smartphones and OBD-II based devices in insurance already work in the way we're proposing. Logs only need to persist for the duration preceding their processing: here, about 42 kilobytes per interval on average. On-chain storage is limited to a risk score value, a string containing the list of violated rules per interval, and a string providing the URL of evidence data off-chain: easily under 1 kilobyte per publication even with many rules. Uploads of encrypted evidence data are similarly reasonable being always less than or equal to the amount of data logged in an interval: no more than a few megabytes per interval even with 100 times the points.\footnote{With the exception of instances when a user includes audio, image, or video data as evidence, though these are not expected to be a constant inclusion like other data.}

\subsubsection{Time}

The practicality of the Smart Auto Insurance system relies on the logging, processing, and publication of data being able to happen actively without a processing backlog. As established, the quantities of data which need to be published and uploaded are certainly reasonable so our main concern is regarding RiskAssessment and data encryption. These are worth highlighting because the data processing involves running $r$ rules against a point at frequency $f$ Hz for the entire interval of $I$ seconds for each data series $s$. This requires a minimum of $\mathcal{O}(rfIs)$ basic operations, and then encrypting each series that contains rules offending points. Our proof of concept finds that these processes take on average 5.36 ms and 1.49 ms, respectively.\footnote{100,000 examples run with a single core on a laptop processor.} While the example only processed four distinct rules on four separate series, this performance suggests we could have four orders of magnitude slower performance without risking falling behind the 300 second RiskAssessment interval. Performance will decrease through use of a slower processor (such as an embedded device) or greater distinct\footnote{The phrasing of \textit{distinct} rules is particular because related rules with shared intermediaries or fewer rules with more complex conditions may process faster.} rules.

\section{Conclusion}
We have presented a system which can, in principle, enable cost-minimized, usage-based, telematic insurance with claims that are easier to process and negotiate through managed data commitments. The Smart Auto Insurance system allows for
controlling the release of sensitive data to aid in both data privacy and fault negotiation endeavors while also serving as a global record for risk rating rules. It does so while introducing minimal computational or storage requirements. While initial assumptions about the validity of input data may seem strong, we maintain that enforcing the highly transactional nature of the Smart Auto Insurance system creates a game in which users must play fairly to be confident they can gain. Future investigation may include more deeply exploring nuances of the game at hand to ensure assumptions about player behavior hold in practical insurance schemes and more rigorously evaluating what rule structures are permissible (i.e. if there are any practical limitations). Moreover, possible extensions to the protocol such as an active driver feedback mechanism or more direct financial incentivization of lower risk driving through OCC approved stablecoin~\cite{Stablecoin:online} collateral/payments may be considered.

\bibliographystyle{plain}
\bibliography{local.bib}

\begin{thebibliography}{10}

\bibitem{2001}
{\em Advanced encryption standard (AES)}.
\newblock Nov 2001.

\bibitem{OBD-II}
Ishak Aris, Mohamad~Fauzi Zakaria, S.~Bashi, and R~Sidek.
\newblock Development of obd-ii driver information system.
\newblock 03 2007.

\bibitem{Arity:online}
Arity.
\newblock Telematics insurance solutions.
\newblock \url{https://www.arity.com/solutions/insurance-solutions/}.

\bibitem{TelematicsGenderDiscrimination}
Mercedes Ayuso, Montserrat Guillen, and Ana Perez-Marin.
\newblock Telematics and gender discrimination: Some usage-based evidence on
  whether men's risk of accidents differs from women's.
\newblock {\em Risks}, 4:10, 04 2016.

\bibitem{IPFS}
Juan Benet.
\newblock {IPFS} - content addressed, versioned, {P2P} file system.
\newblock {\em CoRR}, abs/1407.3561, 2014.

\bibitem{Filecoin}
Juan Benet and Nicola Greco.
\newblock Filecoin: A decentralized storage network.
\newblock {\em Protocol Labs, San Francisco, CA, USA, Tech. Rep.}, 2018.

\bibitem{Craig2020}
Fabrice Benhamouda, Craig Gentry, Sergey Gorbunov, Shai Halevi, Hugo Krawczyk,
  Chengyu Lin, Tal Rabin, and Leonid Reyzin.
\newblock Can a public blockchain keep a secret?
\newblock In Rafael Pass and Krzysztof Pietrzak, editors, {\em Theory of
  Cryptography - 18th International Conference, {TCC} 2020, Durham, NC, USA,
  November 16-19, 2020, Proceedings, Part {I}}, volume 12550 of {\em Lecture
  Notes in Computer Science}, pages 260--290. Springer, 2020.

\bibitem{Allstate:online}
Allstate Corporation.
\newblock Drive-wise.
\newblock \url{https://www.allstate.com/drive-wise.aspx}.

\bibitem{10.1007/978-3-319-07776-5_43}
Jheng-Syu Jhou and Shi-Huang Chen.
\newblock The implementation of obd-ii vehicle diagnosis system integrated with
  cloud computation technology.
\newblock In Jeng-Shyang Pan, Vaclav Snasel, Emilio~S. Corchado, Ajith Abraham,
  and Shyue-Liang Wang, editors, {\em Intelligent Data analysis and its
  Applications, Volume I}, pages 413--420, Cham, 2014. Springer International
  Publishing.

\bibitem{NAIC:online}
Dimitris Karapiperis.
\newblock Usage-based insurance and vehicle telematics: Insurance market and
  regulatory implications.
\newblock {\em NAIC CIPR Study}, 2015.

\bibitem{Stablecoin:online}
Office of~the Comptroller of~the Currency.
\newblock Occ chief counsel's interpretation on national bank and federal
  savings association authority to use independent node verification networks
  and stablecoins for payment activities.
\newblock
  \url{https://www.occ.gov/news-issuances/news-releases/2021/nr-occ-2021-2a.pdf}.

\bibitem{Snapshot:online}
Progressive.
\newblock Snapshot.
\newblock \url{https://www.progressive.com/auto/discounts/snapshot}.

\bibitem{8697833}
P.~R. {Sawant} and Y.~B. {Mane}.
\newblock Design and development of on-board diagnostic (obd) device for cars.
\newblock In {\em 2018 Fourth International Conference on Computing
  Communication Control and Automation (ICCUBEA)}, pages 1--4, 2018.

\bibitem{CasualtyActuarialSociety}
Casualty~Actuarial Society.
\newblock Making the economics of telematics work for insurers.
\newblock {\em Insurance Journal}, 2014.

\bibitem{SmartInsurance}
Jianan Su, Michael Bartholic, Andrew Stange, Ryosuke Ushida, and Shin'ichiro
  Matsuo.
\newblock How to dynamically incentivize sufficient level of iot security.
\newblock In {\em Financial Cryptography and Data Security - {FC} 2020
  International Workshops}, volume 12063 of {\em Lecture Notes in Computer
  Science}, pages 451--465. Springer, 2020.

\end{thebibliography}

\end{document}